\documentclass[preprint, 5p]{elsarticle}

\usepackage{amsmath,amssymb,bm}
\usepackage{lineno}
\usepackage{color}

\usepackage{multirow}
\usepackage{multicol}

\journal{Physics Letters B}

\begin{document}

\begin{frontmatter}

\title{Chiral dynamics in soft triaxial nuclei}

\author[PKU]{B. Li}
\author[PKU]{P.W. Zhao\corref{mycorrespondingauthor}}
\ead{pwzhao@pku.edu.cn}
\author[PKU]{J. Meng}

\address[PKU]{State Key Laboratory of Nuclear Physics and Technology, School of Physics, Peking University, Beijing 100871, China}

\cortext[mycorrespondingauthor]{Corresponding authors}

\begin{abstract}
The chirality in soft triaxial nuclei is investigated for the first time with the time-dependent and tilted axis cranking covariant density functional theories on a three-dimensional space lattice in a microscopic and self-consistent way.
Taking the puzzling chiral nucleus $^{106}$Ag as an example, the experimental energies of the observed nearly degenerate bands are well reproduced without any free parameters beyond the well-defined density functional.
A novel chiral mode in soft triaxial nuclei is newly revealed from the microscopic dynamics of the total angular momentum.
This opens a new research area for the study of chirality, particularly in relation to soft nuclear shapes.
\end{abstract}


\end{frontmatter}

\section{Introduction}
Chirality is a well-known phenomenon in many fields, such as chemistry, biology, molecular physics and material science.
In nuclear physics, chirality was originally suggested by Frauendorf and Meng in triaxial nuclei~\cite{Frauendorf1997Nucl.Phys.A131}, which have three different principal axes of the ellipsoidal density distribution, i.e., the short, intermediate, and long axes.
When a triaxial nucleus rotates around an axis lying outside the three planes spanned by the principal axes, the three angular momentum vectors along the short, intermediate, and long axes in the body-fixed frame could couple to each other in either a left- or right-handed mode.
The two chiral systems are related by the chiral operator $TR(\pi)$ that combines time reversal $T$ and spatial rotation by $\pi$.
The broken chiral symmetry in the body-fixed frame should be restored in the laboratory frame.
This leads to the so-called chiral doublet bands, which consist of a pair of nearly degenerate rotational bands.

It is apparent that the triaxial shape is an essential prerequisite for nuclear chirality.
An ideal picture of nuclear chirality is associated with a rigid triaxial shape, which means that the nuclear surface arising from collective motion of many nucleons has only small fluctuations around the equilibrium shape.
While the study of nuclear chirality is so far focused on the rigid triaxial shape, the chirality in soft triaxial nuclei is poorly known.
However, most triaxial nuclei are ``soft'', which implies large fluctuations around the equilibrium shape for the nuclear surface.
This brings a major challenge to the understanding of nuclear chirality.

So far, evidences of chiral doublet bands have been reported experimentally in $A\sim 80$, 100, 130, and 190 mass regions; see e.g., Refs.~\cite{Starosta2001Phys.Rev.Lett.971,Zhu2003Phys.Rev.Lett.132501,Vaman2004Phys.Rev.Lett.32501,Grodner2006Phys.Rev.Lett.172501,Joshi2007Phys.Rev.Lett.102501,Mukhopadhyay2007Phys.Rev.Lett.172501,Ayangeakaa2013Phys.Rev.Lett.172504,Kuti2014Phys.Rev.Lett.32501,Tonev2014Phys.Rev.Lett.52501,Liu2016Phys.Rev.Lett.112501,Grodner2018Phys.Rev.Lett.22502,Xiong2019At.DataNucl.DataTables193}.
However, the manifestation of chirality in realistic nuclei is usually very  complicated due to the  softness of nuclear shape~\cite{Meng2010JPhysG.37.64025,Meng2016PhysicaScripta53008}.
In particular, there are numerous experimental indications suggesting that triaxial softness may have significant implications for the chiral phenomena; see for examples, Refs.~\cite{Starosta2001Phys.Rev.Lett.971,Vaman2004Phys.Rev.Lett.32501,Tonev2006Phys.Rev.Lett.52501,Petrache2006Phys.Rev.Lett.112502,Joshi2007Phys.Rev.Lett.102501,Ayangeakaa2013Phys.Rev.Lett.172504,Lieder2014Phys.Rev.Lett.202502,Rather2014Phys.Rev.Lett.202503,Liu2016Phys.Rev.Lett.112501}.
One of the most intriguing and elusive chiral candidates with soft triaxiality is $^{106}$Ag, in which a set of strongly coupled doublet bands, crossing each other at spin $I\sim 14\hbar$, were first observed in Ref.~\cite{Joshi2007Phys.Rev.Lett.102501}.
Systematics of chiral doublet bands in the $A\sim 100$ region indicates a strong influence of triaxial softness on the stability of the chiral geometry.
The two pursuant lifetime measurements~\cite{Lieder2014Phys.Rev.Lett.202502,Rather2014Phys.Rev.Lett.202503} have reported consistent data for the $B(E2)$ and $B(M1)$ values but contrasting interpretations.
In particular, a third band lying only slightly higher than the two existing crossing bands in $^{106}$Ag was reported in Ref.~\cite{Lieder2014Phys.Rev.Lett.202502}.
Similar band structure has also been observed in $^{134}$Pr~\cite{Starosta2001Phys.Rev.Lett.971,Tonev2006Phys.Rev.Lett.52501,Timar2011Phys.Rev.C44302}.

Theoretically, the chiral nature of the intriguing three closing bands has been extensively studied with phenomenological models, which however, could lead to distinct interpretations, because they are all adjusted to the data in one way or another with schematic assumptions.
For instance, the nuclear shape is assumed to be rigid in the particle rotor model, which is amongst the most widely used models for nuclear chirality~\cite{Frauendorf1997Nucl.Phys.A131,Koike2004Phys.Rev.Lett.172502,Peng2003Phys.Rev.C44324,Zhang2007Phys.Rev.C44307,Qi2009Phys.Lett.B175,Chen2018Phys.Lett.B744}.
The interacting boson fermion-fermion model introduces a cubic interaction to simulate the triaxiality and its softness
~\cite{Tonev2006Phys.Rev.Lett.52501,Tonev2007Phys.Rev.C44313,Brant2008Phys.Rev.C34301}, while the corresponding strength is unknown and needs to be fitted to data.

A microscopic treatment is thus important to explore the origin of nuclear chirality in soft triaxial nuclei.
The tilted axis cranking (TAC) approach allows one to study nuclear chirality based on a microscopic mean field~\cite{Dimitrov2000Phys.Rev.Lett.5732}.
For a more fundamental investigation, self-consistent methods based on more realistic nuclear interactions have been developed in the framework of nonrelativistic and relativistic density functional theories (DFTs)~\cite{Madokoro2000Phys.Rev.C61301,Olbratowski2004Phys.Rev.Lett.52501,Zhao2017Phys.Lett.B1}.
In particular, the chiral conundrum in $^{106}$Ag was studied with the tilted axis cranking covariant density functional theory (TAC-CDFT)~\cite{Zhao2019Phys.Rev.C54319}, and it showed that the potential energy surface is rather soft in the triaxial direction.

The TAC approach can only describe the lower band of the chiral doublet bands, since it gives either one achiral solution or two degenerate chiral ones while cannot describe the left-right excitation mode.
In view of the recent remarkable progress on the time-dependent covariant density functional theory (TDCDFT) in three-dimensional (3D) lattice space~\cite{Ren2020Phys.Lett.B135194,Ren2020Phys.Rev.C44603}, the dynamics of the chiral nucleus $^{135}$Nd was first studied in Ref.~\cite{Ren2022Phys.Rev.C11301}, and the upper band of the chiral doublet bands was well reproduced by analyzing the dynamics of the so-called \emph{chiral precession}.

In this Letter, the chirality in soft triaxial nuclei is studied for the first time with the state-of-the-art approach, i.e., TAC-CDFT and TDCDFT.
Taking the triaxial soft chiral nucleus $^{106}$Ag as an example, the observed nearly degenerate bands are calculated without any free parameters beyond the well-defined density functional.
The impact of the triaxial softness on nuclear chirality is clearly revealed from the microscopic dynamics of the chiral excitations.

\section{ The lower band of chiral doublet bands}

The starting point of covariant density functional theory (CDFT) is a universal energy density functional~\cite{Ring1996Prog.Part.Nucl.Phys.193,Vretenar2005Phys.Rep.101,Meng2006Prog.Part.Nucl.Phys.470,Niksic2011Prog.Part.Nucl.Phys.519,Meng2015}.
For nuclear chirality, the functional is transformed into a body-fixed frame rotating with a constant angular velocity vector $\bm{\omega}$, which is along an arbitrary direction in space, i.e., the three-dimensional TAC-CDFT~\cite{Zhao2017Phys.Lett.B1}.
The corresponding Kohn-Sham equation for nucleons is a static Dirac equation,
\begin{equation}\label{eq_KS}
  \left[\bm{\alpha}\cdot(\hat{\bm{p}}-\bm{V})+V^0+\beta(m+S)-\bm{\omega}\cdot\hat{\bm{J}}\right]\psi_k(\bm{r}) = \varepsilon_k\psi_k(\bm{r}),
\end{equation}
where $\hat{\bm{J}}$ is the angular momentum, and the relativistic scalar $S(\bm{r})$ and vector $V^\mu(r)$ fields are connected in a self-consistent way to the nucleon densities and current distributions, which are obtained from the Dirac spinors $\psi_k(\bm{r})$.
The complicated interplay between the large Lorentz scalar and vector self-energies~\cite{Ren2020Phys.Rev.C21301} allows the self-consistent treatment of the spin degrees of freedom and the nuclear currents induced by the spatial parts of the vector self-energies, which play an essential role in rotating nuclei; see also Refs.~\cite{Meng2013Front.Phys.55,Meng2015} for details.
The angular velocity $\bm{\omega}$ plays a role of a Lagrange multiplier, and it is connected to the expectation values $\langle\hat{\bm{J}}\rangle$ and the angular momentum quantum number (spin) $I$ by the semiclassical relation $\langle\hat{\bm{J}}\rangle\cdot\langle\hat{\bm{J}}\rangle=I(I+1)$.

In this work, the density functional PC-PK1~\cite{Zhao2010Phys.Rev.C54319} is employed, and the calculations are free of additional parameters.
The Dirac equation \eqref{eq_KS} is solved in a 3D cubic box with the length $22.4$ fm and the mesh size $0.8$ fm along the $x$, $y$, and $z$ axes.
We focus on the chirality in the odd-odd nucleus $^{106}$Ag, which was first reported experimentally in Ref.~\cite{Joshi2007Phys.Rev.Lett.102501} with two crossing bands, and a third close band was reported later in Ref.~\cite{Lieder2014Phys.Rev.Lett.202502}.
According to our previous study~\cite{Zhao2019Phys.Rev.C54319}, it is expected that bands 2 and 3 could be a pair of chiral twin bands based on a four-quasiparticle configuration $\pi g_{9/2}\otimes\nu h_{11/2}(gd)^2$, while band 1 corresponds to a two-quasiparticle configuration $\pi g_{9/2}\otimes\nu h_{11/2}$.
\begin{figure}[!htbp]
\centering
\includegraphics[width=0.45\textwidth]{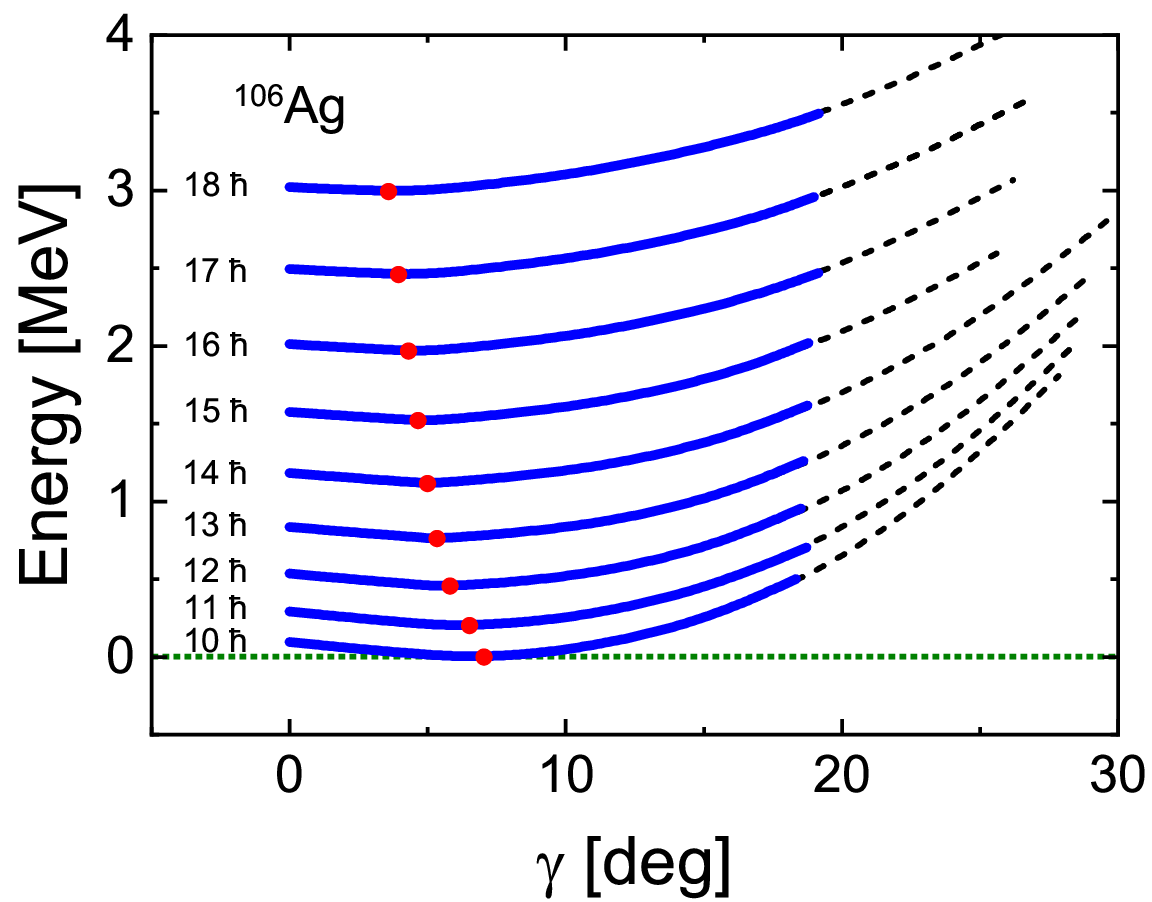}
\caption{(color online). Potential energy curves (dashed line) against the triaxial deformation $\gamma$ for $^{106}$Ag calculated with the TAC-CDFT at each angular momentum with the configuration $\pi g_{9/2}\otimes\nu h_{11/2}(gd)^2$. The solid circles denote the positions of the minimum energies, and the thick lines highlight the region with the corresponding energy rising within $500$ keV against the minima.
}
\label{fig1}
\end{figure}

Figure~\ref{fig1} depicts the potential energy curves against the triaxial deformation $\gamma$ at each angular momentum based on the four-quasiparticle configuration $\pi g_{9/2}\otimes\nu h_{11/2}(gd)^2$.
They are determined by minimizing the total energy at each angular momentum with respect to the $\beta$ deformation for a given value of $\gamma$.
While the triaxial deformation is not large at the energy minima (around $\gamma \simeq 5^\circ$), the potential energy curves are rather soft in the triaxial direction at all angular momenta.
The energy rise against the minimum is in general less than 500 keV when the triaxial deformation reaches a significant value $\gamma \simeq 20^\circ$.
Considering the fact that the four-quasiparticle configuration contains high-$j$ protons and neutrons in the $g_{9/2}$ and $h_{11/2}$ shells, respectively, a chiral excitation mode could thus be generated though deeply influenced by the $\gamma$ softness.
\section{ The higher band of chiral doublet bands}

In order to describe the chiral excitations, the TDCDFT calculations are carried out for the dynamical evolution of the system.
In TDCDFT, the time-dependent Kohn-Sham equation reads~\cite{Ren2020Phys.Lett.B135194,Ren2020Phys.Rev.C44603},
\begin{equation}\label{eq_TDKS}
  i\partial_t\psi_k(\bm{r},t) = [\bm{\alpha\cdot}(\hat{\bm{p}}-\bm{V})+V^0+\beta(m+S)]\psi_k(\bm{r},t),
\end{equation}
where the time-dependent fields $S(\bm{r},t)$ and $V^\mu(\bm{r},t)$ have the same dependence on density and currents as in TAC-CDFT, so there are no additional parameters introduced.
Starting from the TAC-CDFT solutions at each angular momentum, one can obtain the dynamical evolution of the system by means of the time-dependent wave functions $\psi_k(\bm{r},t)$, and the expectation value of the angular momentum is conserved during the time evolution.
As in our previous study of chiral precession~\cite{Ren2022Phys.Rev.C11301}, it is convenient to analyze the chiral dynamics in the body-fixed rotating frame in terms of the tilted angles $\theta_J$ and $\varphi_J$ of the total angular momentum.
Here, $\theta_J$ is the polar angle between the total angular momentum and the long axis, and $\varphi_J$ is the azimuth angle between the projection of the total angular momentum in the short-intermediate plane and the short axis.

\begin{figure}[!htbp]
\centering
\includegraphics[width=0.45\textwidth]{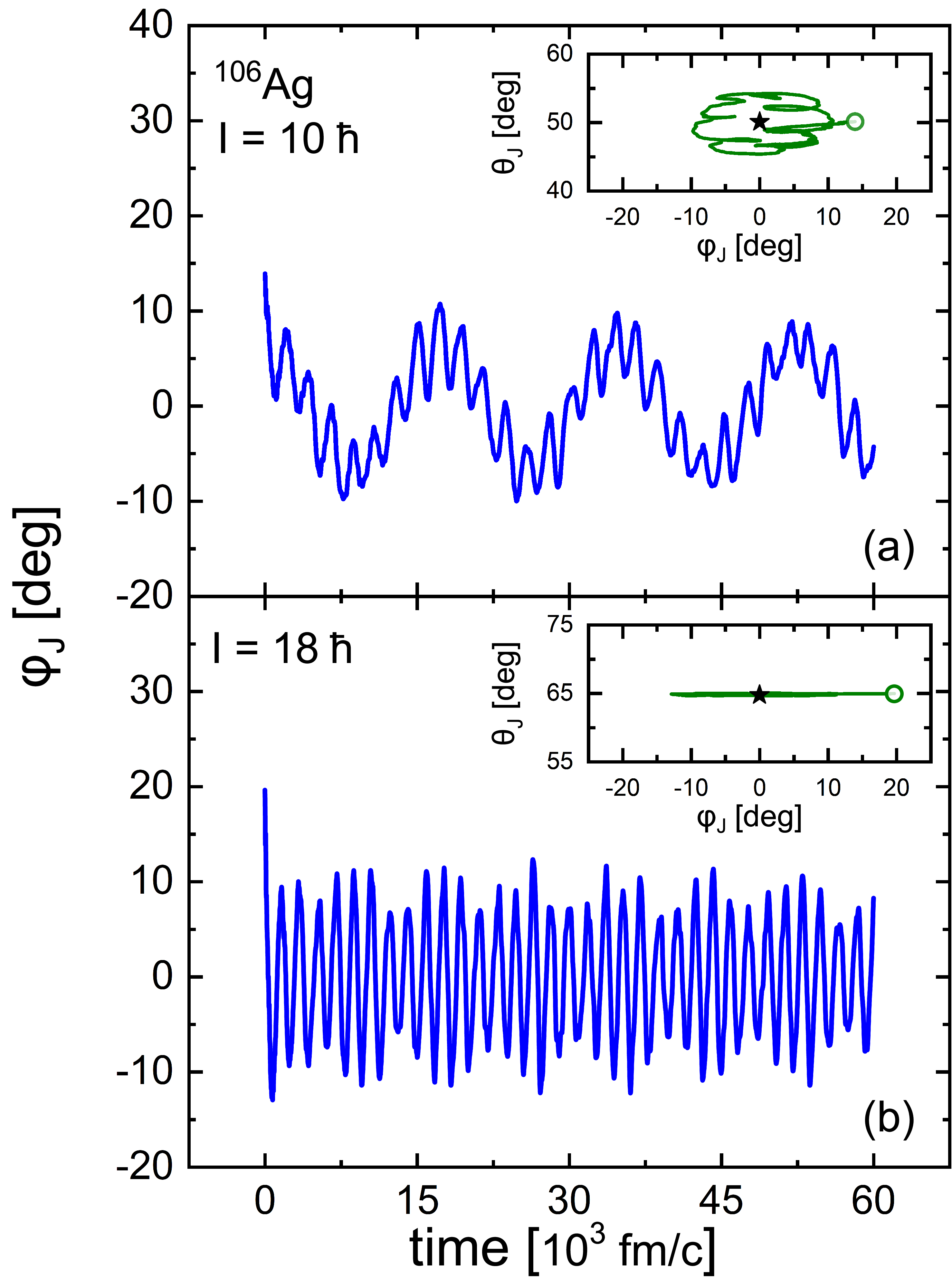}
\caption{(color online). Time evolution of the azimuth angle $\varphi_J$ starting from an initial state taken from the total energy surface in the $(\theta_J, \varphi_J)$ plane at the total angular momentum $I=10\hbar$ (top) and $I=18\hbar$ (bottom).  Insets: Time-dependent trajectories of the tilted angles $(\theta_J, \varphi_J)$. The stars denote the positions of the energy minima for the total energy surface in the $(\theta_J, \varphi_J)$ plane.}
\label{fig2}
\end{figure}

In the inset of Fig.~\ref{fig2}(a), the time-dependent trajectory of the tilted angles $\theta_J$ and $\varphi_J$ is depicted for the total angular momentum $I=10\hbar$.
The initial state is taken from the total energy surface in the $(\theta_J, \varphi_J)$ plane, and it is away from the lowest energy state.
The trajectory can roughly be regarded as an ellipse centered at the location of the minimum energy $\varphi_J = 0$, while it is
significantly distorted with small oscillations in the $\varphi_J$ direction, which are essentially induced by the $\gamma$ softness of $^{106}$Ag.
From the time evolution of $\varphi_J$ as depicted in Fig.~\ref{fig2}(a), one can see even more clearly the two explicit components with distinct frequencies.

One should keep in mind that for nuclei with a rigid shape, the nuclear deformation barely changes with different orientations of the angular momentum, while for a soft nucleus like $^{106}$Ag, the deformation can be easily changed across the energy surface in the $(\theta_J, \varphi_J)$ plane.
As a result, for $^{106}$Ag, starting from an arbitrary state on the energy surface in the $(\theta_J, \varphi_J)$ plane, except for the energy minimum, one could in principle stimulate motions not only in the $(\theta_J, \varphi_J)$ space but also in the deformation space, in particular, in the $\gamma$ direction.
This explains the two explicit components in the chiral dynamics obtained for $^{106}$Ag.
Note that the moderate triaxial deformation of $^{106}$Ag considerably softens the energy surface in the $\varphi_J$ direction, so the precession period is much longer than that for the chiral precession found in $^{135}$Nd~\cite{Ren2022Phys.Rev.C11301}.

Increasing the angular momentum to, for instance, $I=18\hbar$, the lower-frequency mode is invisible, and the time evolution of $\varphi_J$ is dominated by a single frequency harmonic oscillation, as depicted in Fig.~\ref{fig2}(b).
This is indeed due to the fact that the energy surface becomes much softer in the $\varphi_J$ direction and stiffer in the $\theta_J$ direction at higher angular momenta; indicating an extremely long-period mode which actually does not correspond to a real vibration mode.
The soft energy surface is also reflected by the time-dependent trajectory of the tilted angles $\theta_J$ and $\varphi_J$, which is extremely narrowed in the $\theta_J$ direction [see the inset of Fig.~\ref{fig2}(b)].

\begin{figure}[!htbp]
\centering
\includegraphics[width=0.45\textwidth]{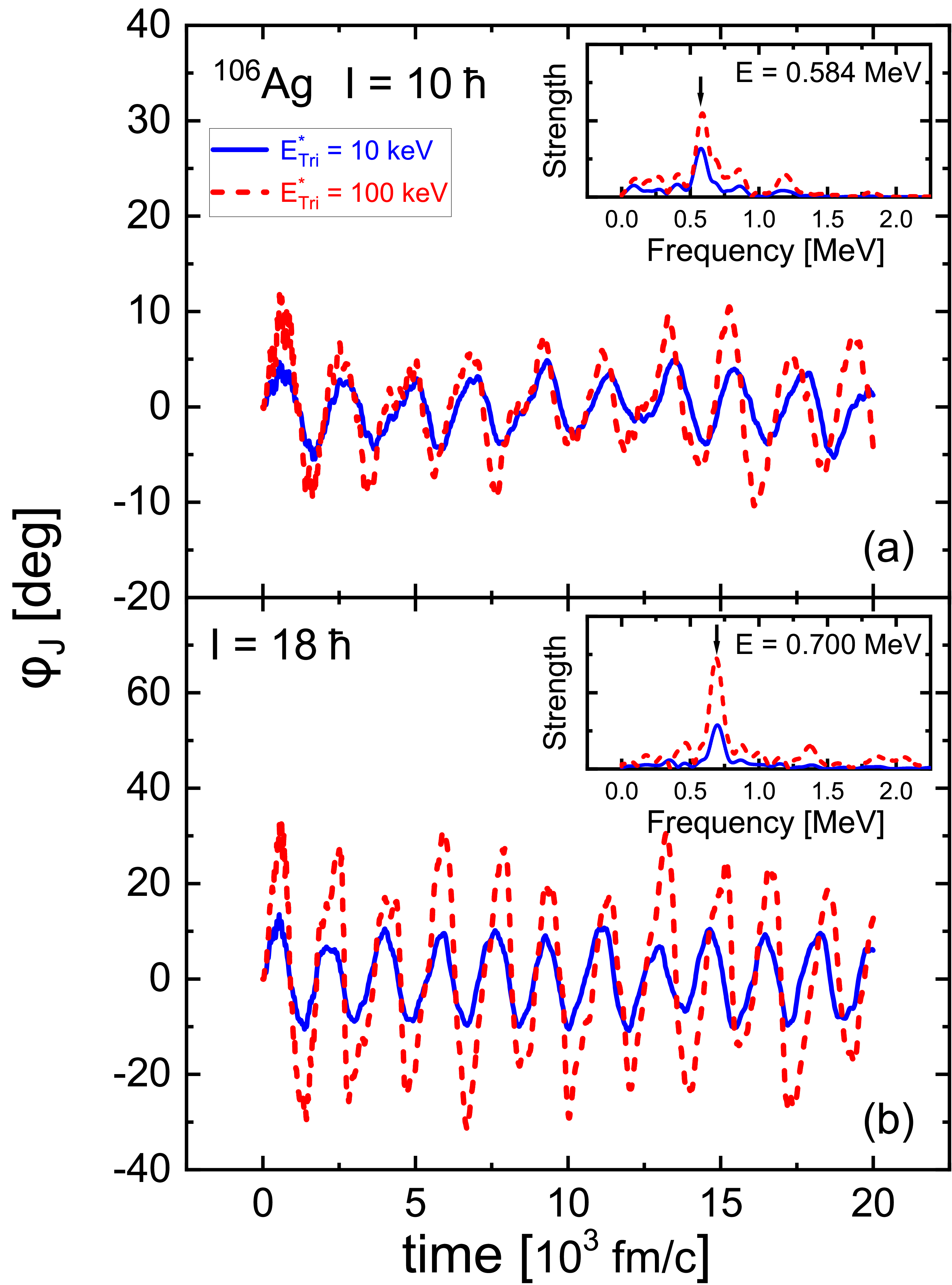}
\caption{(color online). Time evolution of the azimuth angles $\varphi_J$ starting from two initial states which are away from the lowest energy state only in the $\gamma$ direction by 10 keV ($\gamma= 9^\circ$) and 100 keV ($\gamma= 12^\circ$) at the total angular momentum $I=10\hbar$ (top) and $I=18\hbar$ (bottom). Insets: The corresponding spectral powers obtained by Fourier transforms on the time evolution of the azimuth angles, and the arrow indicates the centroid energies.}
\label{fig3}
\end{figure}

The energy of the quantum phonons corresponding to the chiral vibration should be determined by quantizing the semiclassical time-dependent solutions, and it can be easily obtained by extracting the normal-mode frequencies with Fourier transformation in the case of small harmonic oscillations.
From Fig.~\ref{fig2}, one see clearly that the lower-frequency mode plays only a minor role in the chiral dynamics even at $I=10\hbar$ due to the moderate triaxial deformation of $^{106}$Ag.
Moreover, the corresponding phonon energy for the lower-frequency mode would be further suppressed due to the quantum admixture of the two normal modes.
Therefore, here we concentrate on the higher-frequency normal mode, which is mainly associated with the $\gamma$ softness.

To extract the normal-mode frequency with Fourier transformation, a pure higher-frequency normal mode is needed, which can be obtained by performing the TDCDFT calculations starting from certain initial states, i.e., the states differing from the equilibrium state only in the $\gamma$ direction, while the corresponding $\theta_J$, $\varphi_J$,  and $\beta$ values are the same as the equilibrium state.

At $I=10\hbar$, two different initial states are taken as examples, and they are higher in energy than the equilibrium state by 10 keV ($\gamma= 9^\circ$) and 100 keV ($\gamma= 12^\circ$).
The corresponding time evolution of the azimuth angle $\varphi_J$ starting from these two initial states are depicted in Fig.~\ref{fig3}(a).
The initial $\varphi_J$ values are zero for both trajectories, and along the time evolution, they start to oscillate between positive and negative values.
This reflects the chiral nature of the oscillations induced by the $\gamma$ softness.
Moreover, the two trajectories have distinct amplitudes but roughly identical frequency, proving the harmonic nature of the chiral dynamics in such a soft triaxial nucleus.

The harmonic nature of the oscillations can be more clearly seen in the inset of Fig.~\ref{fig3}(a), where the spectral powers obtained by Fourier transforms~\cite{Press1992B,Ren2022Phys.Rev.C11301} on the time evolution of the azimuth angle are depicted.
The height of the spectral power reflects the amplitude of the oscillations, while the frequency is associated with the one-phonon energy due to the chiral motions.
A single main peak is clearly seen for both spectral powers, and the positions of both peaks are identically at 0.584 MeV.
This provides the energy difference for the chiral partner bands at $I=10\hbar$.

Such analysis can be straightforwardly extended to the states at other angular momenta from  $I=11\hbar$ to $I=20\hbar$.
Taking the $I = 18\hbar$ state as an example, in Fig.~\ref{fig3}(b), the time evolution of the azimuth angle $\varphi_J$ is depicted starting from two different initial states, which are away from the equilibrium state only in the $\gamma$ direction.
In general, the behavior of the time evolution at $I = 18\hbar$ is quite similar to that at $I = 10\hbar$, except for the much larger amplitudes, which are associated with the fact that the total energy surface becomes much softer in the $\varphi_J$ direction at higher angular momenta.
The oscillation of the $\varphi_J$ values looks even more harmonic than that at $I = 10\hbar$.
As a result, the spectral powers obtained with Fourier transforms on the time evolution of the azimuth angle, as seen in the inset of Fig.~\ref{fig3}(b), have two peaks at an identical position, i.e., 0.700 MeV.

\begin{figure}[!htbp]
\centering
\includegraphics[width=0.45\textwidth]{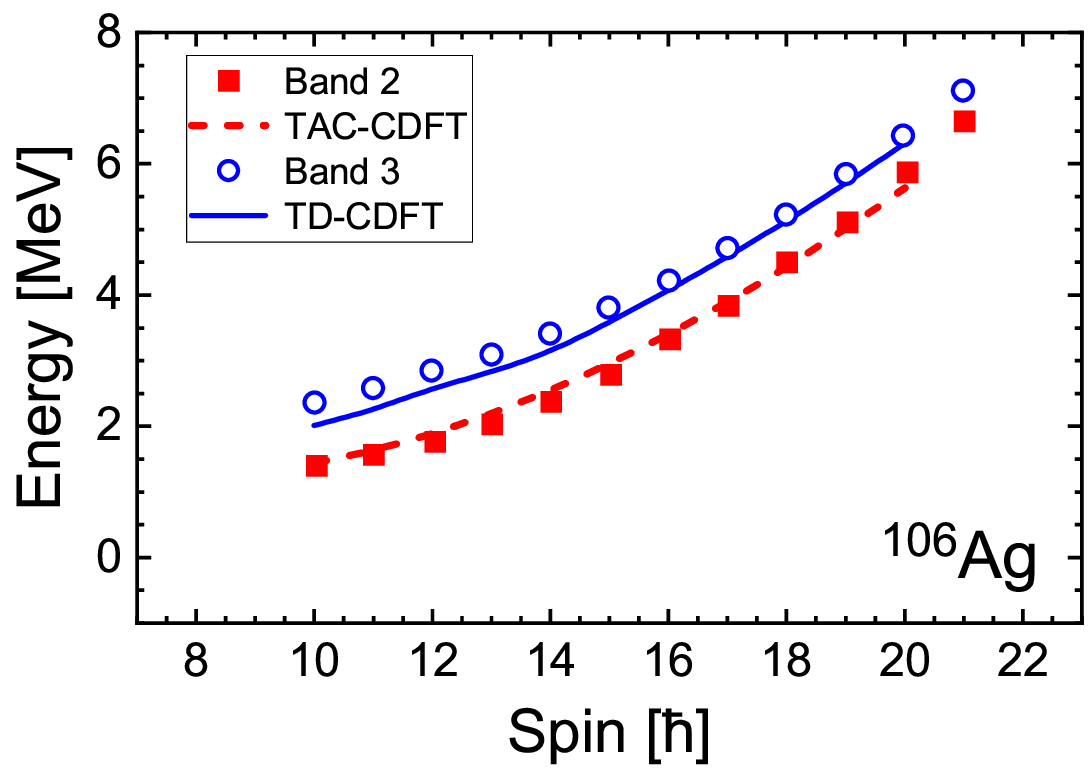}
\caption{(Color online). Calculated energies (solid and dashed lines) for chiral doublet bands in $^{106}$Ag built on the four-quasiparticle configuration $\pi g_{9/2}\otimes\nu h_{11/2}(gd)^2$  in comparison with data (solid and open symbols)~\cite{Lieder2014Phys.Rev.Lett.202502}.
  The dashed line represents the TAC-CDFT results, and the solid one represents the results including the chiral excitation energies obtained with the TDCDFT calculations.}
\label{fig5}
\end{figure}

Finally, by combining the calculated results with both TAC-CDFT and TDCDFT at each spin, one could provide a microscopic description for the partner bands observed in the triaxial soft nucleus $^{106}$Ag.
In Fig.~\ref{fig5}, the calculated energy spectra for the partner bands built on the four-quasiparticle configuration $\pi g_{9/2}\otimes\nu h_{11/2}(gd)^2$ are depicted in comparison with data~\cite{Lieder2014Phys.Rev.Lett.202502}.
For the lower band, the excitation energies are originated from planar rotations and, thus, are calculated with the TAC-CDFT calculations.
Based on the TAC-CDFT solutions at each spin, the TDCDFT calculations provide the corresponding excitation energies due to the chiral motion.
Consequently, the upper band can be built by adding up the TAC-CDFT energies and the corresponding chiral excitation energy at every spin.
It can be seen that the experimental energies are well reproduced.
This explains the puzzling chiral structure in $^{106}$Ag and provides the first microscopic description for the chiral doublet bands in a triaxial soft nucleus.

\section{Summary}
In summary, for the first time, the chirality in soft triaxial nuclei has been investigated in the microscopic time-dependent covariant density functional theory, and the initial state at a given angular momentum is taken from the calculations within the three-dimensional tilted axis cranking covariant density functional theory.
Taking the puzzling chiral nucleus $^{106}$Ag as an example, the experimental energies of the observed nearly degenerate bands are well reproduced without any free parameters beyond the well-defined density functional.
From the microscopic dynamics of the total angular momentum in the body-fixed frame, it is found that the precession dynamics, appeared in rigid chiral nuclei, could be remarkably suppressed in soft chiral nuclei.
Instead, a novel chiral mode in soft triaxial nuclei occurs and it is responsible for the chiral excitations observed in $^{106}$Ag.
This not only explains the puzzling chiral structure in $^{106}$Ag, but also opens a new research area for the study of chirality, particularly in relation to soft nuclear shapes.

\section*{Conflict of interest}
The authors declare that they have no known competing financial
interests or personal relationships that could have appeared to influence
the work reported in this paper.

\section*{Acknowledgments}
We thank Zhengxue Ren for fruitful discussions.
This work was partly supported by the National Natural Science Foundation of China (Grants No. 12070131001, No. 11935003, No. 11975031, and No. 12141501),
and the High-performance Computing Platform of Peking University.


\end{document}